\documentclass[journal]{IEEEtran}

\newcommand{\dd}{{\rm d}}

\usepackage{graphicx}
\usepackage{amsmath}

\newcommand{\wid}{4.5cm}
\newcommand{\widd}{-6mm}

\errorstopmode

\begin{document}

\title{Non-Euclidean cloaking for light waves}

\author{Tom\'a\v{s} Tyc, Huanyang Chen, Che Ting Chan, and  Ulf Leonhardt
\thanks{T. Tyc is with Institute of Theoretical Physics and Astrophysics,
Masaryk University, Kotlarska 2, 61137 Brno, Czech Republic}
\thanks{H. Chen and C. T. Chan are with Department of Physics, The Hong Kong
  University of Science and Technology, Clear Water Bay, Hong Kong, People's
  Republic of China}
\thanks{U. Leonhardt is with School of Physics and Astronomy, University of
  St~Andrews, North Haugh, St Andrews KY16 9SS, UK} }


\maketitle

\begin{abstract}

Non-Euclidean geometry combined with transformation optics has recently led to
the proposal of an invisibility cloak that avoids optical singularities and
therefore can work, in principle, in a broad band of the spectrum [U. Leonhardt
  and T.  Tyc, Science 323, 110 (2009)]. Such a cloak is perfect in the limit
of geometrical optics, but not in wave optics. Here we analyze, both
analytically and numerically, full wave propagation in non-Euclidean
cloaking. We show that the cloaking device performs remarkably well even in a
regime beyond geometrical optics where the device is comparable in size with
the wavelength. In particular, the cloak is nearly perfect for a spectrum of
frequencies that are related to spherical harmonics. We also show that for
increasing wavenumber the device works increasingly better, approaching perfect
behavior in the limit of geometrical optics.

\end{abstract}

\begin{IEEEkeywords}
Electromagnetic cloaking, Transformation optics, Wave propagation
\end{IEEEkeywords}

\IEEEpeerreviewmaketitle

\section{Introduction}

\IEEEPARstart{E}{lectromagnetic} cloaking is a modern and rapidly developing
area of optics.  The main objective of cloaking is to encapsulate an object
into a cloak, a carefully designed material with special optical properties,
such that the object becomes invisible. This is achieved by bending light rays
(or wave fronts) such that they flow smoothly around the object and after
passing it, restore their original direction, creating the illusion that the
space is empty.  Several types of cloaks have been proposed
\cite{Pen06,UlfSci06,Ulf09, Che07, Lai09, Li08,Smo09}, some of which have been
realized experimentally \cite{Sch06, Liu09, Val09, Gab09}. All are based on
transformation optics \cite{Pen06,UlfSci06,Ulf09,Dol61,Gre03,GREE,overviewUlf},
the idea that optical media effectively transforms the geometry for light, and
all, except \cite{Smo09}, require metamaterials \cite{Shalaev, Milton,SPW,
SLW,Sarychev} for their practical realization.  Originally, two different
approaches have been used to propose an invisibility cloak.

The proposal by J.B. Pendry \cite{Pen06} used a geometric transformation of
Euclidean space to expand a single point into a region of finite volume. Based
on ideas of transformation optics, a metamaterial was designed that would lead
light around this region, making it invisible. A drawback of this proposal was
the necessity of infinite light speed at the boundary of the invisible region,
which prevents such a device from working in a broad band of the spectrum. The
proposal was realized experimentally for microwaves \cite{Sch06} and provided
an inspiration for other similar ideas.

The proposal by U. Leonhardt \cite{UlfSci06} also applied transformation
optics, but used methods of complex analysis for designing an invisibility
cloak that could be made with an isotropic medium.  In particular, it took
advantage of multi-valued analytic functions and the rich structure of Riemann
sheets and branch cuts in the complex plane. An important ingredient was a
refractive index distribution with focusing properties such that a ray that
left the main Riemann sheet could return back to it and continue in the
original direction. However, the speed of light in this proposal diverged as
well, so it could not work broadband either.

The paradigm of cloaking was changed by the proposal of U. Leonhardt and T. Tyc
\cite{Ulf09}. They showed that the necessity of infinite light speed can be
avoided by further developing the ideas of \cite{UlfSci06} and proposed a
cloaking device based on a combination of transformation optics and
non-Euclidean geometry. All optical parameters of the material have
non-singular values in the whole space and therefore nothing prevents them, in
principle, from working broadband. This proposal may turn invisibility into a
practical technology.  In contrast to Euclidean cloaking \cite{Pen06}, however,
non-Euclidean cloaking works perfectly only in the limit of geometrical
optics. For wave optics, imperfections emerge due to diffraction and phase
delays. These imperfections do not matter, though, in most practical
situations, because they become negligible if the size of the device is much
larger than the wavelength. At the same time, it would be very useful to know
what the character of the imperfections is and how the cloaking device behaves
if the validity condition for geometrical optics is not satisfied. This is the
purpose of this paper. We present an analysis of wave propagation in the 2D
version of the cloaking device \cite{Ulf09} and provide computer simulations of
wave propagation in the cloak.

The paper is organized as follows. In Sec.~\ref{TO} we explain the idea of
transformation optics and in Sec.~\ref{cloak} we describe the non-Euclidean
cloaking device. In Sec.~\ref{material} we calculate the optical properties of
the material of the cloak and in Sec.~\ref{waves} we analyze theoretically wave
propagation in the cloak. In Sec.\ref{numerical} we support our theoretical
analysis by numerical simulations of the wave propagation in the cloak and we
conclude in Sec.~\ref{conclusion}.

\section{Transformation optics}
\label{TO}

In this section we briefly and intuitively explain the idea of transformation
optics, referring to the papers \cite{GREE,overviewUlf} for the full technical
details and techniques of calculating the material properties of the optical
medium.  For simplicity, we will consider the limit of geometrical optics and
ignore the polarization of light.

Transformation optics relies on the fact that an optical medium effectively
changes the geometry of space perceived by light.  What determines light
propagation is not the geometric path length, but rather the optical path
length, which should be, according to Fermat's principle, stationary (usually
minimal, but in some cases it is maximal) for the actual path of the ray. The
optical path element is $n$-times larger than the geometric path element, where
$n$ is the refractive index. Hence, the medium creates an effective ``optical
metric tensor'' for light that equals $n^2$-times the ``geometric metric
tensor'', and light rays then propagate along geodesics with respect to this
optical metric tensor.  In case of an anisotropic refractive index (such as in
crystals or metamaterials), the transformation of the metric is just slightly
more complicated.

Conversely, suppose that light propagates in an empty space (so-called {\em
virtual space}) $\boldsymbol{R}=(X,Y,Z)$ with a uniform refractive index where
ray trajectories are straight lines. If we map this space by some function $f$
to a space $\boldsymbol{r}=(x,y,z)$ (so-called {\em physical space}) such that
$\boldsymbol{r}=f(\boldsymbol{R})$, the geometric metric is not preserved in
general, because $\dd x^2+\dd y^2+\dd z^2\not=\dd X^2+\dd Y^2+\dd
Z^2$. However, filling physical space with a suitable optical medium, we can
compensate for this and make the optical metrics in virtual and physical spaces
equal. Then in physical space light will follow the images $f(\boldsymbol{l})$
of straight lines $\boldsymbol{l}$ of virtual space. By choosing a suitable
transformation $f$, one can achieve various interesting effects such as
invisibility \cite{Pen06} or perfect focusing to a single point of light rays
coming from all directions \cite{Ulf08}.  In general the mapping $f$ is not
conformal (i.e., it does not isotropically stretch or squeeze an infinitesimal
neighborhood of a given point) and therefore the resulting refractive index in
physical space is not isotropic.

Remarkably, transformation optics works much more generally than just in the
simplified situation we have just described. In particular, it is valid not
just in the limit of geometrical optics, but it works equally well for full
electromagnetic waves \cite{GREE}.  If the electric permittivity and magnetic
permeability are appropriately chosen \cite{GREE}, Maxwell's equations in
physical space are equivalent to Maxwell's equations in empty virtual space. It
turns out \cite{GREE} that the tensors of relative electric permittivity and
relative magnetic permeability must be proportional to each other to avoid
birefringence and therefore to ensure that both light polarizations behave the
same way in the medium.  Ideally, the two tensors are equal, which gives
impedance matching with the vacuum \cite{Jac99} and hence causes no reflection
at the interfaces of the device.

\section{Non-Euclidean cloaking device}
\label{cloak}

Next we describe the non-Euclidean cloaking device~\cite{Ulf09} using pictures
and try to give some intuition for the geometric transformation used.  We will
leave the mathematical details for the next section; a full description of the
transformation including all calculations can be found in \cite{SOM}, the
Supporting Online Material of \cite{Ulf09}.

The most distinctive feature of the cloaking device \cite{Ulf09} is that
virtual space is not just the simple Euclidean space as in other proposals but
contains a non-Euclidean part. We will focus on the two-dimensional version of
the cloak here; the generalization to three dimensions is relatively
straightforward and has been described in \cite{Ulf09,SOM}.

\begin{figure}[htb]
\begin{center}
\includegraphics[width=8cm]{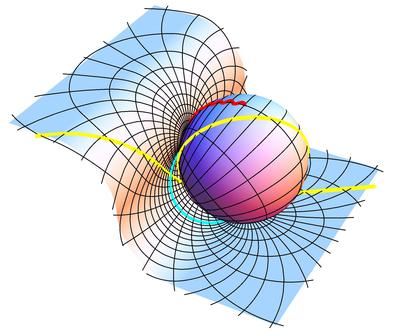}
\end{center}
\caption{Virtual space consists of a sphere and a plane that are connected
  along a line segment -- branch cut.}
\label{plane_and_sphere}
\end{figure}

Virtual space consists of two parts: (i) a Euclidean plane $P$ and (ii) a
2-sphere $S$ (surface of a 3D ball).  The two parts are connected along a
branch cut, a 1D line that on the sphere corresponds to one quarter of its
equator while in the plane it corresponds to a straight line segment. The best
way to visualize this is to imagine that the plane is partly wrapped around the
sphere so that $P$ and $S$ touch each other along $l$ (see
Fig.~\ref{plane_and_sphere}).

A light ray propagating in the plane $P$ that hits the branch cut (at the point
$X$, say) transfers to the sphere $S$. There it propagates along a geodesic,
which is a great circle, and returns to the point $X$ where it transfers back
to the plane $P$ and continues in the original direction. In this way, the
sphere is invisible and the only feature that can possibly disclose its
presence is a time delay of the rays that have entered the sphere.

\begin{figure}[htb]
\begin{center}
\includegraphics[width=6cm]{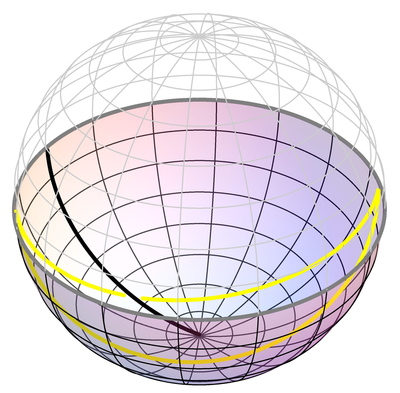}
\end{center}
\caption{When a mirror (gray thick line) is placed along a great circle
 (equator in the picture), any light ray (yellow line) still returns to its
 original position at the branch cut (black line) where it entered the sphere
 but never visits the northern hemisphere; it therefore becomes invisible.}
\label{mirror}
\end{figure}

To create an invisible region where an object could be hidden, one can take
advantage of the high symmetry of the sphere. Placing a mirror (see
Fig.~\ref{mirror}) along a suitably chosen great circle on the sphere, the
light ray, after two reflections, still returns to the point $X$ but completely
avoids the hemisphere hidden behind the mirror; anything placed there becomes
invisible.

\begin{figure}[htb]
\begin{center}
\includegraphics[width=6cm]{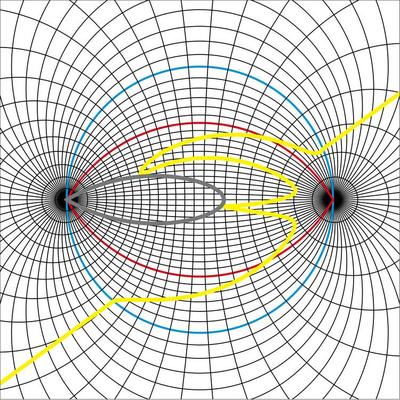}
\end{center}
\caption{Physical space is a plane; the sphere $S$ of virtual space is mapped
  to the inside of the red curve while the plane $P$ of virtual space is mapped
  to the outside of the curve. The hidden hemisphere is mapped to the inside of
  the gray curve which becomes the invisible region of the device.}
\label{physical_plane}
\end{figure}

The physical space of the device is a Euclidean plane filled with an optical
medium.  The map from virtual to physical space is based on bipolar coordinates
and maps both the sphere $S$ and the plane $P$ of virtual space to regions of
the physical plane (see Fig.~\ref{physical_plane}).  The map contains no
singularities, i.e., no points where space would be infinitely stretched or
squeezed. Therefore the speed of light in the medium is finite and nonzero
everywhere, which is a great advantage compared to Euclidean cloaks. The
invisible region is the image of the hidden hemisphere of virtual space.

\section{Material properties of the cloak}
\label{material}

In this paper we focus on a two-dimensional version of the non-Euclidean
invisibility cloak that has been described in the previous section.  To adjust
the two-dimensional situation to real propagation of electromagnetic waves that
are governed by Maxwell equations and are therefore intrinsically
three-dimensional, we do the following construction.  We perform the above
described geometric transformation from two-dimensional virtual space to the
plane $xy$ of physical space and then we add an extra dimension $z$, making
physical space three-dimensional.  In calculating the line elements in this 3D
physical space, we then have to add the term $\dd z^2$ to the line elements
corresponding to the two-dimensional situation.  We also assume that the
wavevector of the incoming wave is perpendicular to the $z$-axis and hence the
phase and amplitude of the wave do not depend on $z$. This makes the problem
effectively two-dimensional, just as required. 

To describe the Euclidean part of the cloak, we use bipolar cylindrical
coordinates $(\tau,\sigma,z)$ in physical space that are related to the
Cartesian $(x,y,z)$ as
\begin{equation}
x = \frac{a \, \sinh \tau}{\cosh \tau - \cos \sigma} \,\, , \quad
y = \frac{a \, \sin \sigma}{\cosh \tau - \cos \sigma} \,\,,\quad z=z.
\label{bipolar}
\end{equation}
Here the parameter $a$ determines the size of the cloak.  Analogous relations
hold in virtual space where $x,y$ and $\sigma$ are replaced by $x',y'$ and
$\sigma'$, respectively.  The mapping between the two spaces is given in the
bipolar cylindrical coordinates by the following relation between $\sigma$ and
$\sigma'$:
\begin{align}
\sigma' &= \sigma \quad\quad\quad\quad
\quad\quad\quad\quad\quad\quad\quad
  \mbox{for}\quad  |\sigma|\leq \displaystyle
  \frac{\pi}{2}\\
\sigma'& =\left(\frac{4\sigma^2}\pi-3|\sigma|+\pi\right)\rm sgn \,\sigma
  \quad  \,\mbox{for}\quad  \frac{\pi}{2} < |\sigma| \leq \frac{3\pi}4
\label{sigma'sigma}\end{align}
and the coordinates $\tau,z$ coincide in the two spaces.  The line element $\dd
s^2=\dd x'^2+\dd y'^2+\dd z^2$ in virtual space is equal to
\begin{equation}
  \dd s^2=\frac{a^2}{(\cosh\tau-\cos\sigma')^2}\,(\dd\tau^2+\dd\sigma'^2)+\dd z^2.
\end{equation}
The permittivity and permeability tensors for this region can be then
calculated using the same general method \cite{GREE} as in \cite{SOM} with the
result 
\begin{equation}
  \varepsilon_i^j=\mu_i^j={\rm diag}\,\left[\frac{\dd\sigma}{\dd\sigma'},\frac{\dd\sigma'}{\dd\sigma},\frac{(\cosh\tau-\cos\sigma)}{(\cosh\tau-\cos\sigma')}\frac{\dd\sigma'}{\dd\sigma}\right] \,.
\end{equation}

The mapping between the non-Euclidean part of virtual space and the
corresponding part of physical space is more involved.  We will not reproduce
here all the corresponding formulas of \cite{SOM}, but rather mention just the
most important steps.
One starts with a sphere of radius
\begin{equation}
r = \frac{4}{\pi}\,a
\end{equation}
(see Eq.~(S43) of \cite{SOM})
parameterized by spherical coordinates -- the latitudal angle $\theta$ and the
longitudinal angle that is denoted by $\sigma'$.  This is not yet the sphere of
of virtual space, but is related to it by a suitable M\"obius transformation
(Eq.~(S40) of \cite{SOM}) represented on the sphere via stereographic
projection (Eqs.~(S37)--(S39) of \cite{SOM}).  Then the coordinates
$(\sigma',\theta)$ are mapped to the bipolar coordinates $(\sigma, \tau)$ of
physical space as follows. The mapping between $\sigma'$ and $\sigma$ is given
by Eq.~(\ref{sigma'sigma}) where $\sigma'$ is set to one of the intervals
$[-2\pi,-\pi]$ and $[\pi,2\pi]$ by adding or subtracting $2\pi$.  The
corresponding intervals of $\sigma$ are then $[-\pi,-3\pi/4]$ and
$[3\pi/4,\pi]$. The mapping between $\theta$ and $\tau$ can be derived by
combining Eqs. (S41) and (S46) of \cite{SOM}. This yields after some
calculations
\begin{equation}
  \theta=2\arctan(t-1+\sqrt{t^2+1}) 
\end{equation}
with
\begin{equation}
  t=\tan\left[\frac\pi4\left(\frac{\sinh\tau}{\cosh\tau+1}\right)+\pi/4\right].
\end{equation}
The line element of the non-Euclidean part of virtual space is 
\begin{multline}
\dd s^2=\frac{16a^2}{\pi^2}\,\left[\frac{1+\cot^2(\theta/2)}{1-2\cos\sigma'\cot(\theta/2)+2\cot^2(\theta/2)}\right]^2 \\
\times (\dd\theta^2+\sin^2\theta\,\dd\sigma'^2)+\dd z^2.
\label{lineelement}\end{multline}
This line element is slightly different from the line element (S47) in
\cite{SOM}.  In fact, the latter turned out to be wrong, because it is the line
element on the sphere $(X'Y'Z')$ as described in section ``Non-Euclidean
branch'' of \cite{SOM}, while the correct line element should be measured on
the sphere $(XYZ)$. The factor in the rectangular parentheses in
Eq.~(\ref{lineelement}) expresses the scaling of the line element when going
from one sphere to the other, as can be calculated from the transformation
(S40) of \cite{SOM} between the two spheres.

After some algebra, Eq.~(\ref{lineelement}) can be brought into the form
\begin{multline}
  \dd s^2=\left[\frac{2(t^2+1+t\sqrt{t^2+1})}{1+(t-1+\sqrt{t^2+1})^2}\right]^2\frac{u^2}{(\cosh\tau+1)^2}\dd\tau^2\\
 +\frac{16a^2}{\pi^2}\left[\frac{2(t-1+\sqrt{t^2+1})}{1+(t-1+\sqrt{t^2+1})^2}\right]^2 u^2\,\dd\sigma'^2+\dd z^2
\end{multline}  
with 
\begin{equation}
  u=\frac{(t-1+\sqrt{t^2+1})^2+1}{(t-1+\sqrt{t^2+1})^2-2(t-1+\sqrt{t^2+1})\cos\sigma'+2} \,.
\end{equation}

Following again the general method \cite{GREE}, we finally obtain the
permittivity and permeability tensors for the region $3\pi/4\le|\sigma|\le\pi$
in bipolar cylindrical coordinates:
\begin{equation}
  \varepsilon_i^j=\mu_i^j={\rm diag}\,(
  \varepsilon_\sigma,\varepsilon_\tau,\varepsilon_z) ,
\end{equation}
with
\begin{align}
  \varepsilon_\sigma&
 =\frac\pi4\frac{t^2+1+t\sqrt{t^2+1}}{t-1+\sqrt{t^2+1}}\,\frac{1}{\cosh\tau+1}\,\frac{\dd\sigma}{\dd\sigma'},\\
  \varepsilon_\tau&
=\frac1{\varepsilon_\sigma}\\\nonumber
  \varepsilon_z&=\frac{16}\pi\,\frac{(t-1+\sqrt{t^2+1})(t^2+1+t\sqrt{t^2+1})}{[1+(t-1+\sqrt{t^2+1})^2]^2}\\&\qquad\times\frac{u^2(\cosh\tau-\cos\sigma)^2}{\cosh\tau+1}\,\frac{\dd\sigma'}{\dd\sigma} .
\end{align}

\begin{figure}[htb]
\begin{center}
\includegraphics[width=6cm]{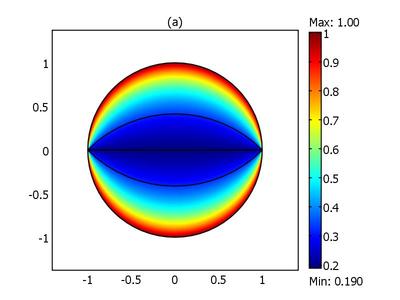}\\
\includegraphics[width=6cm]{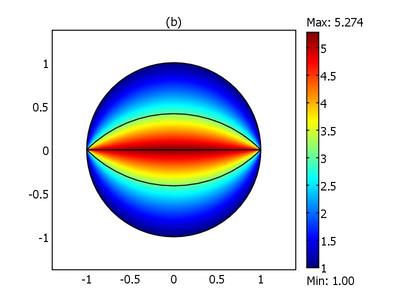}\\
\includegraphics[width=6cm]{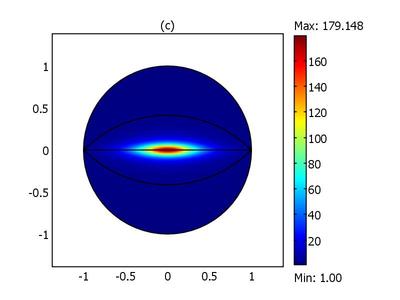}
\end{center}
\caption{The values of $\mu_\sigma$ (a), $\mu_\tau$ (b) and $\varepsilon_z$ (c)
  in the region of the cloaking device; the black line marks the border between
  the Euclidean and non-Euclidean region. All the values are non-singular.}
\label{tensor_values}
\end{figure}

Figure \ref{tensor_values} shows the values of
$\mu_\sigma,\mu_\tau$ and $\varepsilon_z$ in the cloaking region, respectively.
We picked these material parameters, because in our simulations we used 
transverse electric (TE) waves whose electric field is parallel with the
$z$-axis, and in this case the material parameters that influence the wave
propagation are only $\mu_\sigma$, $\mu_\tau$ and $\varepsilon_z$.
We found numerically that
$\mu_\sigma\in[0.190,1]$, $\mu_\tau\in[1, 5.274]$ and
$\varepsilon_z\in [1, 179.15]$.  Clearly, all the values avoid zero and
infinity and hence non-Euclidean cloaking requires no optical singularity of
the material.

To create the invisible region, a mirror is placed in virtual space along a
great circle on the sphere. Since an ideal mirror can be
described by a material with infinite permittivity, it retains this property
upon a geometric transformation and remains an ideal mirror.  Therefore in
physical space we simply place the mirror along the curve that is the image of
the great circle.

There is also another option for creating an invisible region apart from using
the mirror, namely expanding a line of physical space that is not crossed by
any ray. We will not consider this option in this paper.

\section{Wave propagation in the invisibility cloak}
\label{waves}

First we will focus on the situation where there is no invisibility region yet;
there is no mirror and the whole sphere of virtual space is accessible to
light. Then we proceed to the situation with the mirror.

To see what happens when a light wave propagates in the cloak, we need to
understand the behavior of the waves in the non-Euclidean region of the
device. This region is an image of a sphere in virtual space with uniform
refractive index.  Wave propagation on such a sphere is different from that in
an Euclidean space, and the solutions of the wave (or, equivalently, Helmholtz)
equation are described by the spherical harmonics $Y_{lm}(\theta,\varphi)$.
The situation here is still more complicated, though. The sphere in virtual
space is not closed but is ``cut'' along the branch cut, and the two sides of
the branch cut are connected to different parts of the plane of virtual
space. Therefore the amplitude and phase of the wave on the two sides of the
branch cut need not be the same as would be the case of a normal, connected
sphere.  The result is that there can be also other waves propagating on this
sphere beside the spherical harmonics.

It can happen, though, that the incoming wave in the plane is tuned in such a
way that that the wave on the sphere will naturally have the same amplitude and
phase at any two points adjacent across the branch cut, so it will behave as a
wave on a normal, uncut sphere. Evidently, this can only happen for waves that
are resonant on the sphere, i.e., for the modes described by spherical
harmonics.

To find the corresponding frequencies, we recall that the spherical harmonics
$Y_{lm}(\theta,\varphi)$ are solutions of the Helmholtz equation $\Delta
u+k^2u=0$ on the sphere. Here $l$ is a non-negative integer and $m$ is an
integer satisfying $-l\le m\le l$. The corresponding quantized wavevectors
satisfy the relation
\begin{equation}
  k_l=\frac{\sqrt{l(l+1)}}r,
\label{k_l}\end{equation}
where $r$ is the radius of the sphere, and depend on $l$ but not on $m$. This
degeneracy with respect to $m$ is related to the high symmetry of the sphere.
The corresponding frequencies and wavelengths are then
\begin{equation}
  \omega_l=\frac cr\,\sqrt{l(l+1)}, \qquad  \lambda_l=\frac{2\pi
    r}{\sqrt{l(l+1)}}. 
\label{omega,lambda}\end{equation}
We therefore expect that the cloak will behave perfectly for light waves with
wavevectors and frequencies given by Eqs.~(\ref{k_l})
and~(\ref{omega,lambda}). 

Another way of seeing this is the following. A spherical harmonic describes a
wave running on the sphere. If we make a loop on the sphere and return to the
same place, the value of $Y_{lm}$ returns to its original value. In this sense
we can say that a wave described by a spherical harmonic interferes
constructively with the version of itself that has traveled around the sphere.
And this is exactly what happens to a wave that has entered the sphere via the
branch cut -- it travels on the sphere, is delayed, and then leaves via the
other side of the branch cut for the plane again. If such a wave should
interfere constructively with the wave that has not entered the sphere at all,
the time delay must be equal to a multiple of the period of the wave. Since
this happens to waves with wavevectors and frequencies given by
Eqs.~(\ref{k_l}) and~(\ref{omega,lambda}), these will be the waves for which
the cloak will work perfectly.

For waves whose wavelengths do not satisfy the condition~(\ref{omega,lambda}),
the interference of the waves that entered the sphere with those that did not
will not be constructive and therefore we can expect that the device will not
work perfectly, disturbing the waves.

Note that the constructive interference condition~(\ref{omega,lambda}) is
different from what one might expect by the following (wrong) argument: The
waves that enter the sphere will interfere with those that did not and the
result of this interference will depend on the time delay corresponding to
propagation around the sphere. At first sight, it might seem that this time
delay is simply equal to the length of the equator of the sphere $2\pi r$
divided by speed of light $c$. This would suggest that the device will perform
best if this delay is an integer multiple of the period of the light wave
$2\pi/\omega$, which would then yield the wavelength of light equal to
$\lambda={2\pi r}/m$ with $m$ integer.  However, this is different from the
correct condition~(\ref{omega,lambda}). It is the curvature of the sphere that
modifies wave propagation compared to a flat, Euclidean space and makes such a
simple argument invalid.

Next we consider the full version of the non-Euclidean cloak including the
mirror.  In the limit of geometrical optics, the mirror should make no
difference, because a ray entering the sphere of virtual space at some point
returns to the same point just as if no mirror were present.  Moreover, even
from the semi-geometrical optics point of view there should be no change, as
the ray picks up a total phase of $2\pi$: one $\pi$ for each of the two
reflections from the mirror.

However, looking at the situation with the mirror fully wave-optically, there
should be some change caused by the mirror. Indeed, the mirror imposes a
boundary condition on the wave, in particular, the electric field must be zero
at the mirror surface.  A more detailed analysis shows that instead of $2l+1$
linearly independent solutions $Y_{lm}(\theta,\varphi)$ of the Helmholtz
equation for a given $l$, there are just $l$ independent solutions that vanish
on a given great circle. For example, if the great circle is chosen as the
equator (corresponding to $\theta=\pi/2$), then these solutions will be the
$Y_{lm}$ themselves, but just those with $m$ having different parity than $l$.
Therefore the waves propagating in the mirror cloak do not have such a freedom
as they would if no mirror were present, because of the boundary condition or,
equivalently, because of the restricted set of modes. We can then expect that
the mirror cloak will not work perfectly even for those wavelengths satisfying
the condition~(\ref{omega,lambda}).

\section{Numerical simulations}
\label{numerical}

\begin{figure}
\begin{center}
\includegraphics[width=\wid]{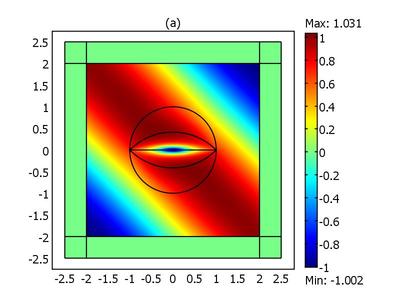} \hspace{\widd}
\includegraphics[width=\wid]{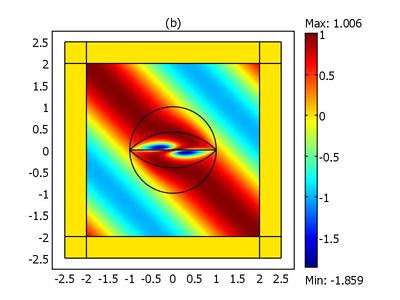}\\
\includegraphics[width=\wid]{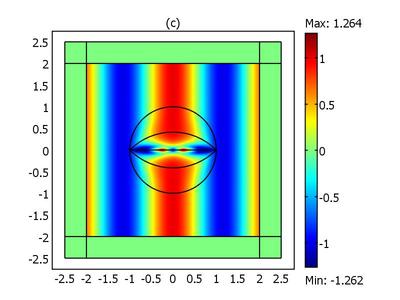}\hspace{\widd}
\includegraphics[width=\wid]{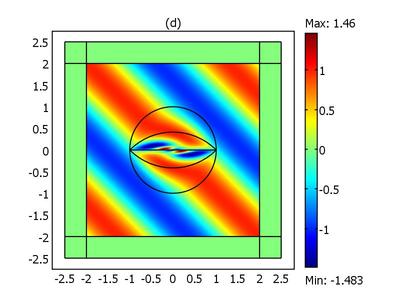}\\
\includegraphics[width=\wid]{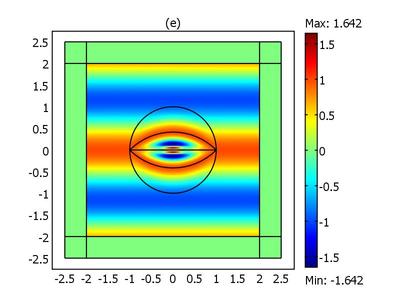}\hspace{\widd}
\includegraphics[width=\wid]{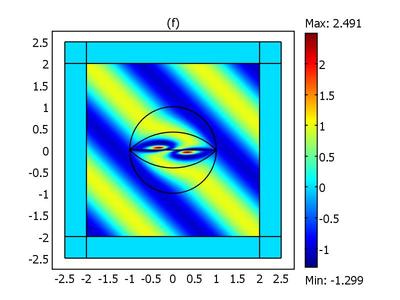}
\end{center}
\caption{Wave propagation in the cloaking device for $l=1$ (a), $l=2$ (b),
 $l=3$ with various angles of incidence (c, d, e), and for $l=4$ (f).  Due
 to constructive interference the waves are not disturbed.}
\label{integer_l_no_mirror}
\end{figure}

Now we proceed to numerical simulations of wave propagation in the cloaking
device to see whether they confirm the above considerations.  Figure
\ref{integer_l_no_mirror} shows the propagation of a wave incident on the cloak
with wavelength satisfying Eq.~(\ref{omega,lambda}) with several different
values of $l$ and several angles of impact. We see that the wave is not
disturbed by the cloak and propagates as if no cloak were present at all.

\begin{figure}
\begin{center}
\includegraphics[width=\wid]{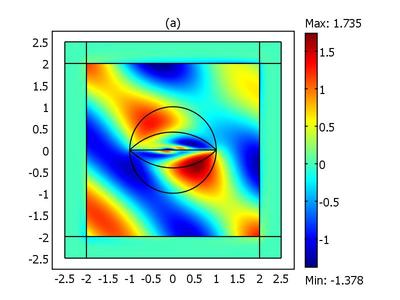} \hspace{\widd}
\includegraphics[width=\wid]{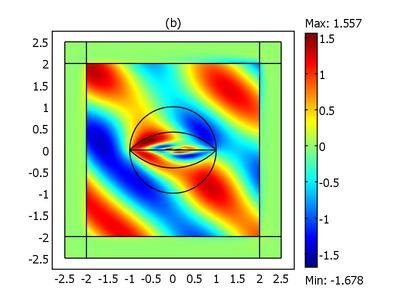}\\
\includegraphics[width=\wid]{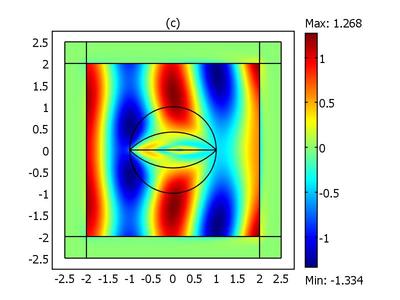}\hspace{\widd}
\includegraphics[width=\wid]{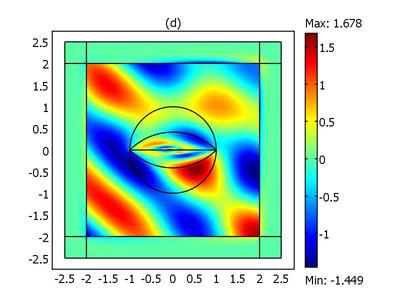}\\
\includegraphics[width=\wid]{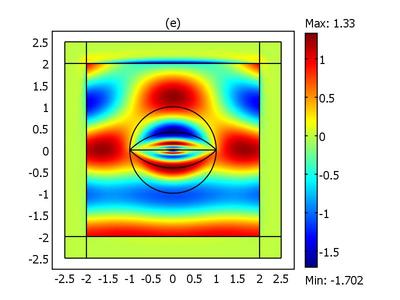}\hspace{\widd}
\includegraphics[width=\wid]{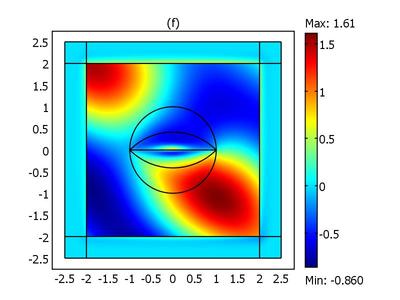}
\end{center}
\caption{Wave propagation at the cloaking device with the
condition~(\ref{omega,lambda}) not satisfied for integer $l$. The values of $l$
are 2.8 (a), 3.2 (b), 3.5 (c, d, e) with various angled of incidence and 1.5
(f).  The device does not work perfectly due to destructive (or partially
destructive) interference of the waves that did and that did not enter the
non-Euclidean part of the device. We see that for half-integer values of $l$
the phase shift of the waves that entered the non-Euclidean part of the device
is approximately $\pi$, which is obvious especially in the pictures (d) and
(e).}
\label{non-integer_l_no_mirror}
\end{figure}

Furthermore, Fig.~\ref{non-integer_l_no_mirror} shows the propagation of waves
that do not satisfy Eq.~(\ref{omega,lambda}) with integer $l$.  To compare the
two cases, we still use, however, the number $l$ as in
Eq.~(\ref{omega,lambda}), but we allow also for its non-integer values.  We see
from Fig.~\ref{non-integer_l_no_mirror} that for non-integer $l$ the waves are
disturbed, just as we predicted. The perturbation is strongest for half-integer
values of $l$, because they are most distant from the resonant integer values.
Figure 6 clearly shows the time lag of the waves propagating through the
cloaking structure, resulting in phase shifts that are strongest for
half-integer $l$.  Nevertheless, the emerging phase fronts are still parallel
to the incoming fronts, which indicates that images are faithfully
transmitted. However, due to the time lag, two wavefront dislocations are
formed, as Fig. 6 also shows. They would cause image distortions at the edges
of the non-Euclidean region of the device.

\begin{figure}
\begin{center}
\includegraphics[width=\wid]{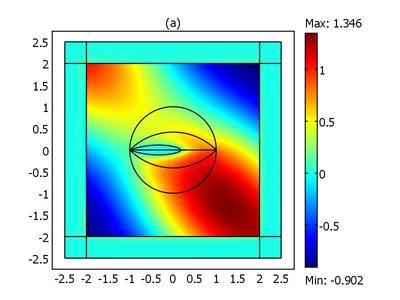} \hspace{\widd}
\includegraphics[width=\wid]{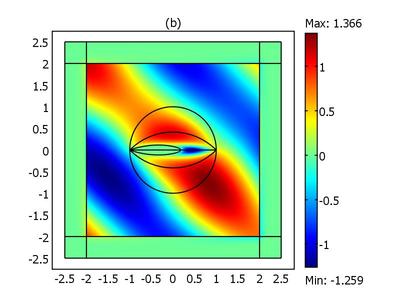}\\
\includegraphics[width=\wid]{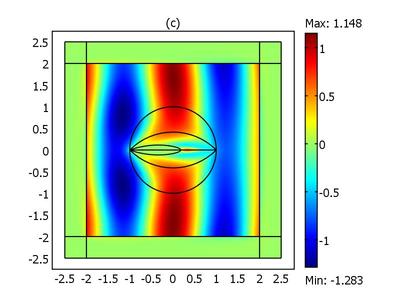}\hspace{\widd}
\includegraphics[width=\wid]{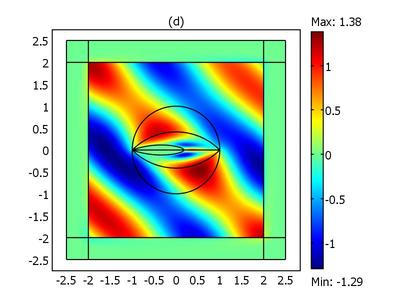}\\
\includegraphics[width=\wid]{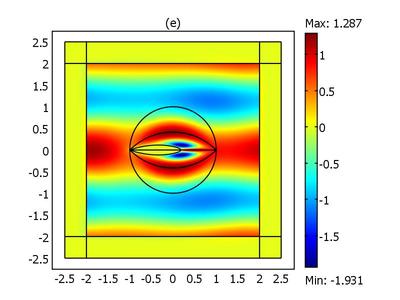}\hspace{\widd}
\includegraphics[width=\wid]{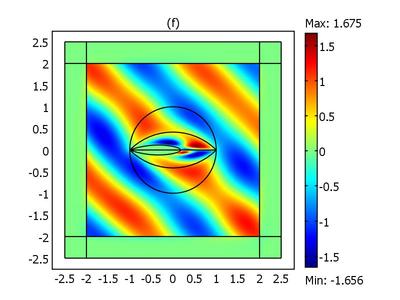}
\end{center}
\caption{Same as Fig.~\ref{integer_l_no_mirror}, that is, $l=1$ (a), $l=2$ (b),
 $l=3$ (c, d, e), $l=4$ (f) but with the mirror included. The mirror is the
 ellipse-like object that encapsulates the invisible region.  Clearly, electric
 field vanishes in this region.  Due to the restricted set of modes the
 performance of the device is no more perfect, but the imperfection is weak. }
\label{integer_l_mirror}
\end{figure}

Next we consider the complete cloaking device with mirror.  Figure
\ref{integer_l_mirror} shows the behavior of waves with integer $l$. Although
there is some disturbance of the waves, the perturbation is quite modest.  This
is quite remarkable if we take into account that the non-Euclidean cloak has
originally been proposed for the regime of geometrical optics, and there was
absolutely no ambition that it works for wavelengths comparable with the size
of the device.  Here we see that the cloak can perform very well also outside
of geometrical optics, and large wavelengths do not seem to be a problem.
Furthermore, Fig. 8 shows that the mirror does not affect the propagation of
waves with non-integer values of $l$. The wave distortions are very similar to
the case without the mirror.


Last we look at the behavior for large values of $l$. Unfortunately, the
simulations became soon very difficult and even on a powerful computer
converged just up to $l=20$. We therefore could not numerically test the
behavior of the device for larger values. Figure~\ref{large_l} shows the case
$l=20$ and $l=19.5$ without and with the mirror. We see that while for $l=20$
the device works nearly perfectly (we think that the imperfections are caused
by numerical errors), for $l=19.5$ the behavior is almost perfect, apart from
interference effects on the edges. Extrapolating this for larger $l$, we can
conclude that the cloaking device should indeed work increasingly well.

\begin{figure}
\begin{center}
\includegraphics[width=\wid]{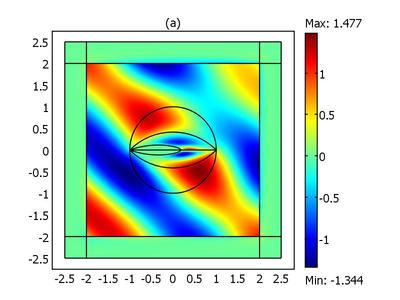} \hspace{\widd}
\includegraphics[width=\wid]{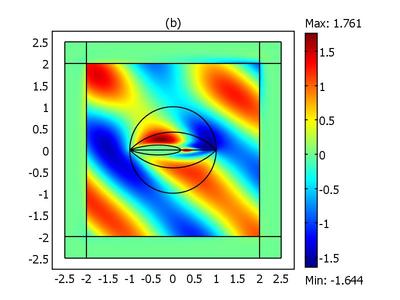}\\
\includegraphics[width=\wid]{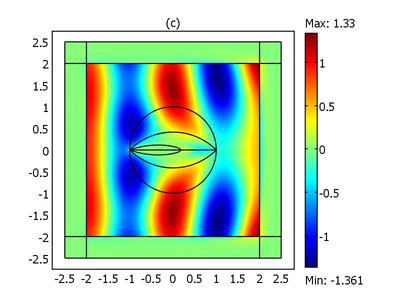}\hspace{\widd}
\includegraphics[width=\wid]{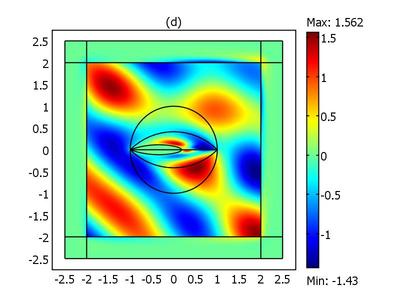}\\
\includegraphics[width=\wid]{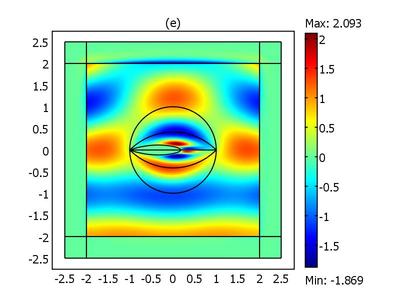}\hspace{\widd}
\includegraphics[width=\wid]{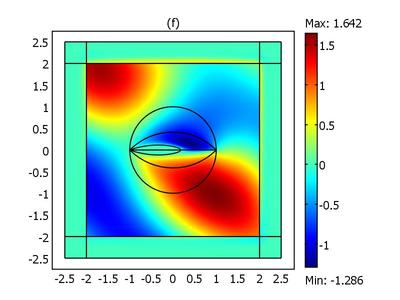}
\end{center}
\caption{Same as Fig.~\ref{non-integer_l_no_mirror}, that is, $l=2.8$ (a),
  $l=3.2$ (b), $l=3.5$ (c, d, e) and $l=1.5$ (f), but with the mirror
 included. The restricted set of modes causes a slightly different performance
 of the device compared to the situation without the mirror.}
\label{non-integer_l_mirror}
\end{figure}

\begin{figure}
\begin{center}
\includegraphics[width=\wid]{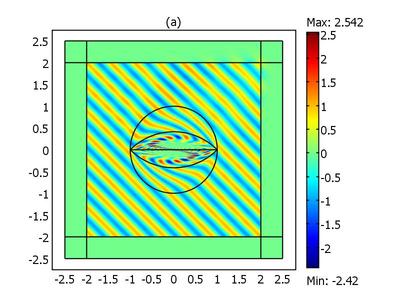} \hspace{\widd}
\includegraphics[width=\wid]{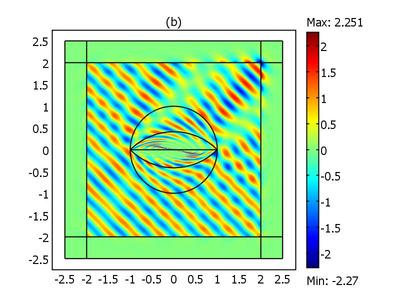}\\
\includegraphics[width=\wid]{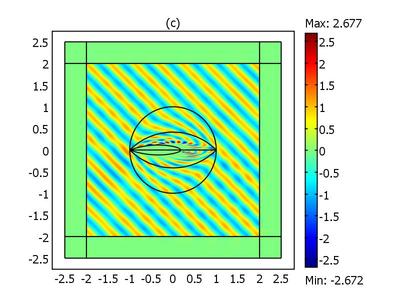}\hspace{\widd}
\includegraphics[width=\wid]{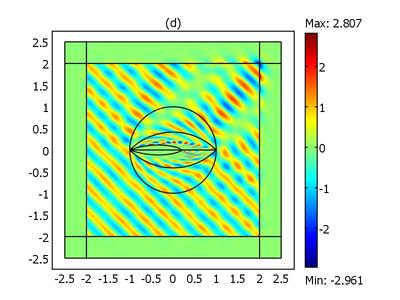}\\
\end{center}
\caption{Wave propagation in the cloak without the mirror (first line) and with
it (second line) for $l=20$ (a), (c) and $l=19.5$ (b), (d).  The device now
performs quite well even for the half-integer $l$.}
\label{large_l}
\end{figure}

\section{Conclusion}
\label{conclusion}

In conclusion, we have analyzed wave propagation in the two-dimensional version
of the non-Euclidean cloaking device~\cite{Ulf09}. We have seen that although
the device works perfectly in the geometrical optics limit, there are
imperfections due to interference effects if full wave description is
employed. In the situation where there is no hidden region, there exist
wavelengths of light for which the device is absolutely perfect. This results
from constructive interference between the waves that entered the non-Euclidean
part of the cloak and those that did not.  This constructive interference
occurs for wavelengths that resonate on the non-Euclidean part of the cloak and
correspond to frequencies of spherical harmonics. For other wavelengths the
interference is not fully constructive and the waves leaving the device are
disturbed --- the device is not perfect for light waves.  The perturbations are
caused by the time lag in the device and result in wavefront dislocations.

Although adding the mirror and creating an invisible region causes the cloak to
cease working perfectly for the resonant wavelengths, the imperfection is very
weak.  The performance of the device is much better than
originally expected \cite{Ulf09}.  When the wavelength gets smaller, 
the performance of the cloaking device gradually improves, becoming
perfect in the limit of geometrical optics.

The fact that the non-Euclidean invisibility cloak does not work perfectly for
light waves is by far outweighed by the absence of optical singularities and
the potential of working broadband. Although non-Euclidean cloaking has been
designed for wavelengths much smaller than the size of the device, which is
also the case of most practical situations, it is useful to understand the
behavior of the device outside this limit, as we have tried to do in this
paper. This might help designing other types of non-Euclidean cloaks that could
possibly work even for wavelengths comparable with the size of the device.

\section{Acknowledgments}

We acknowledge support of the grants MSM0021622409 and MSM0021622419, and a
 Royal Society Wolfson Research Merit Award.



\begin{biography}[{\includegraphics[width=1in,height=1.25in,clip,keepaspectratio]{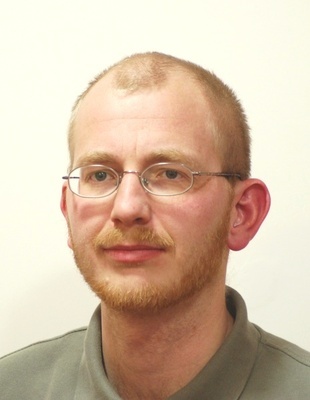}}]
{Tom\'a\v{s} Tyc}
was born in Brno, Czechoslovakia (now Czech Republic) on February 7th, 1973.
He received his Master degree and Ph.D. in theoretical physics from Masaryk
University in Brno in 1996 and 1999, respectively. In his research Tom\'a\v{s}
Tyc initially investigated correlations in free electron beams, later he
switched to quantum optics and quantum information theory. In 2007 he focused
on optics, in particular on theory of invisibile cloaks and transformation
optics. He obtained habilitation (Associate professorship) from Masaryk
University in 2006. He was a Research Fellow at University of Vienna (2000),
Macquarie University Sydney (2001, 2002), University of Calgary (2004), and
University of St Andrews (2007, 2008). Beside his research interests, he likes
to popularize science and makes shows on physics of everyday life for general
public.
\end{biography}

\begin{biography}[{\includegraphics[width=1in,height=1.25in,clip,keepaspectratio]{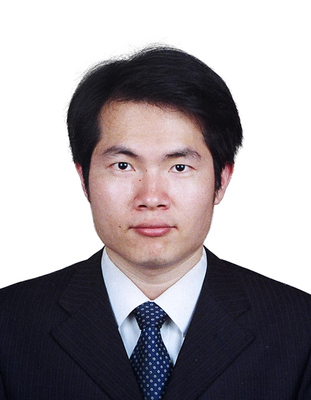}}]
{Huanyang Chen}
was born in Fujian, China, on August 5, 1983. He received his B.Sc. and
Ph.D. in physics from Shanghai Jiao Tong University, Shanghai, China, in 2005
and in 2008.  He was a Research Assistant and a Post-Doctoral Fellow in the
group of Professor Che Ting Chan at Department of Physics, the Hong Kong
University of Science and Technology, Hong Kong, China, from 2006 to 2009. He
will join Suzhou University, Jiangsu, China after September 2009. His research
interests include photonic band gaps, negative index materials, metamaterial
design and transformation optics. He has published several original papers on
transformation optics and cloaking.

\end{biography}

\begin{biography}[{\includegraphics[width=1in,height=1.25in,clip,keepaspectratio]{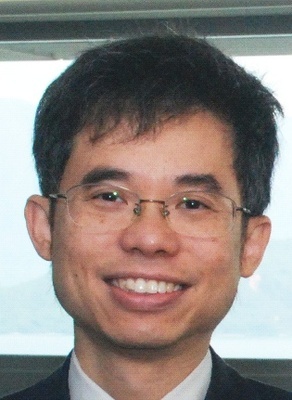}}]
{Che Ting Chan} received his BS degree in Physics from the University of Hong
   Kong in 1980 and his PhD degree from UC Berkeley in 1985. He is currently a
   Chair Professor of Physics at HKUST, and the director of William Mong
   Institute of Nano Science and Technology (WMINST). Dr. Chan's research
   interest has been in the area of photonic crystals, metamaterials and
   nano-materials. Dr. Chan is a Fellow of the American Physical Society.
\end{biography}
  
\begin{biography}[{\includegraphics[width=1in,height=1.25in,clip,keepaspectratio]{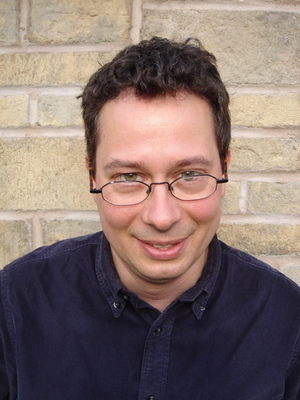}}]
{Ulf Leonhardt}
was born in Schlema, in former East Germany, on October 9th 1965. He studied at
Friedrich-Schiller University Jena, Germany, at Moscow State University,
Russia, and at Humboldt University Berlin, Germany. He received the Diploma in
Physics from Friedrich-Schiller University in 1990 and the PhD in Theoretical
Physics from Humboldt University in 1993. Ulf Leonhardt was a research
associate at the Max Planck Research Group Nonclassical Radiation in Berlin
1994-1995, a visiting scholar at the Oregon Center for Optics in Eugene,
Oregon, 1995-1996, a Habilitation Fellow of the German Research Council at the
University of Ulm, Germany, 1996-1998, and a Feodor-Lynen and Göran-Gustafsson
Fellow at the Royal Institute of Technology in Stockholm, Sweden,
1998-2000. Since April 2000 he is the Chair in Theoretical Physics at the
University of St Andrews, Scotland. In 2008 he was a Visiting Professor at the
National University of Singapore. Ulf Leonhardt is the first from former East
Germany to win the Otto Hahn Award of the Max Planck Society. For his PhD
thesis he received the Tiburtius Prize of the Senate of Berlin. In 2006
Scientific American listed him among the top 50 policy business and research
leaders for his work on invisibility devices. In 2008 he received a Royal
Society Wolfson Research Merit Award. He is a Fellow of the Institute of
Physics and of the Royal Society of Edinburgh.
\end{biography}

\end{document}